\documentclass[aps,pra,twocolumn,groupedaddress,floatfix,showpacs]{revtex4}

\usepackage{amsmath,amsfonts,amssymb,graphics,graphicx,epsfig,color}
\begin{document}
\title{Ground--state properties of trapped Bose--Fermi \\ 
       mixtures: role of exchange--correlation}
\author{Alexander P. Albus$^{1}$, Fabrizio Illuminati$^{2}$, 
and Martin Wilkens$^{1}$}
\affiliation{\mbox{}$^{1}$Institut f\"{u}r Physik, 
Universit\"{a}t Potsdam, 
D--14469 Potsdam, Germany\\
\mbox{}$^{2}$Dipartimento di Fisica, Universit\`{a} di Salerno, 
and Istituto Nazionale per la Fisica della Materia, I--84081 Baronissi
(SA), Italy}

\date{\today}

\begin{abstract}
We introduce density-functional-theory for inhomogeneous
Bose--Fermi mixtures, derive the associated
Kohn-Sham equations, and determine the exchange--correlation 
energy in local density approximation. 
We solve numerically the Kohn--Sham system and
determine the boson and fermion density distributions
and the ground--state energy of a trapped, dilute
mixture beyond mean--field approximation.
The importance of the corrections due
to exchange--correlation is discussed by  
comparison with current experiments;
in particular, we investigate the effect of
of the repulsive potential energy contribution 
due to exchange--correlation on the stability
of the mixture against collapse.

\end{abstract}


\maketitle

\section{Introduction}
The achievement of Bose-Einstein condensation 
in trapped, dilute alkali gases~\cite{BEC}
has stimulated a rapidly growing 
activity in the field of ultracold,
degenerate quantum gases, aimed at a better 
understanding of fundamental aspects of quantum theory. 
In particular, recent experimental progresses have
opened the way to the fascinating prospect of realizing
a BCS transition to superfluidity in ultracold, trapped
Fermi gases.

Magnetically trapped fermions interact very weakly, as their
spins are polarized in the direction of the 
trapping magnetic field, so that  
fermion-fermion $s$-wave scattering is prevented
by the Pauli principle.
Cooling of the fermions to quantum degeneracy can then
be efficiently achieved by mixing them
with ultracold bosons.
After the process of sympathetic cooling, 
the final phase of the system is 
a quantum degenerate Bose-Fermi mixture. 
Indeed, such a system has been recently realized
experimentally~\cite{TueStr01,SchKha01,HadSta02,RoaRib02}.

On the theoretical side, dilute Bose-Fermi mixtures have been 
studied both in homogeneous and confined geometries. 
For homogeneous systems, recent work has addressed the
problem of stability and phase separation~\cite{VivPet00};
the effect of boson-fermion interactions on the 
dynamics~\cite{Meystre}; and the BCS transition induced on 
the fermions by the boson-fermion interactions~\cite{Stoof}. 
The first correction to the ground-state energy 
beyond mean--field approximation has been determined
analytically for homogeneous systems~\cite{Noi}.
This exchange--correlation term can be used for trapped
systems in local density approximation, i.e. when
the interaction length scales are much smaller
than the characteristic sizes of the trapping potentials. 
This condition is naturally met in the current experiments.  
Recent numerical work~\cite{Giorgini} confirms the analytical 
findings in the corresponding regime for homogeneous systems.

For trapped systems the theory has been developed in
mean--field approximation 
to determine the boson and fermion density 
profiles at zero temperature~\cite{Mol98}, and
the related properties of stability against
phase separation and collapse~\cite{Roth}.
A mean field approach has been also employed 
to calculate the critical temperature of Bose-Einstein
condensation in a trapped mixture~\cite{Tcrit}.
However, a description beyond mean field is 
needed either when the interaction parameters 
are large, or to gain a very 
precise knowledge of the density profiles and the
related properties of stability.
In the present work we determine 
the ground--state energy and the boson and fermion 
density profiles to second order in the boson--fermion
scattering length for harmonically
trapped Bose--Fermi mixtures at zero temperature,
and determine the modification, due to the resulting
exchange--correlation energy, of the mean-field
predictions. 

The plan of the paper is the following.
In Section \ref{Theory} we briefly show how to apply
Density Functional Theory (DFT)~\cite{DreGro90}
to inhomogeneous boson-fermion systems, and we
determine the exchange-correlation energy functional
via local density approximation (LDA) 
on the ground-state energy functional
of homogeneous mixtures beyond mean field 
obtained in Ref.~\cite{Noi}. 
In Section \ref{Results} we provide the numerical
solution of the coupled, nonlinear Kohn--Sham
equations for the boson and fermion density
distributions, and we determine the importance
of the corrections due to exchange--correlation
by comparing our results with 
current experiments. In Section \ref{Stability}
we discuss the effect of the 
exchange--correlation energy term on the
phase diagram of the mixture, especially
regarding the onset of collapse for 
mixtures with attractive boson--fermion
interaction.

\section{Theory}
\label{Theory}
 
We begin by considering a inhomogeneous, dilute system of
interacting bosons and spin-polarized fermions
with two-body interactions in $s$-wave scattering
approximation, so that the interparticle potentials
are $U_{BB}(|{\bf r}-{\bf r}'|)=g_{BB}\delta({\bf r}-{\bf r}')$, 
$U_{FF}(|{\bf r}-{\bf r}'|)=0$,
and $U_{BF}(|{\bf r}-{\bf r}'|)=g_{BF}\delta({\bf r}-{\bf r}')$.
The boson-boson coupling is $g_{BB}=4\pi\hbar^{2}a_{BB}/m_{B}$,
where $a_{BB}$ is the boson-boson $s$-wave scattering length,
and $m_{B}$ is the boson mass. The boson-fermion coupling reads 
$g_{BF}=2\pi\hbar^{2}a_{BF}/m_{R}$, where $a_{BF}$ is the
boson-fermion $s$-wave scattering length, 
and $m_{R} = m_{B}m_{F}/(m_{B} + m_{F})$ is the reduced
mass ($m_{F}$ is the fermion mass).
The full Hamiltonian reads:
\begin{equation}
\hat{H}=\hat{T}_B+\hat{T}_F+\hat{V}_B
+\hat{V}_F+\hat{W}_{BB}+\hat{W}_{BF},
\end{equation}
\noindent where
\begin{equation}
\hat{T}_{B}=
-{\displaystyle \int } d{\bf r}\hat{\Phi}^\dagger({\bf r})
\frac{\hbar^2{\bf \nabla}^2}{2m_B}\hat{\Phi}({\bf r}); \,
\hat{V}_{B}=\int d{\bf r} 
\hat{\Phi}^\dagger({\bf r})V_B \hat{\Phi}({\bf r}),
\nonumber
\end{equation}
\begin{equation}
\hat{T}_{F}=
-{\displaystyle \int } d{\bf r}\hat{\Psi}^\dagger({\bf r})
\frac{\hbar^2{\bf \nabla}^2}{2m_F}\hat{\Psi}({\bf r}); \,
\hat{V}_{F}=\int d{\bf r}
\hat{\Psi}^\dagger({\bf r})V_F \hat{\Psi}({\bf r}),
\nonumber
\end{equation}
\begin{equation}
\hat{W}_{BB}=\frac{1}{2}
\int \int d{\bf r} d{\bf r}' \hat{\Phi}^\dagger({\bf r}) 
\hat{\Phi}^\dagger({\bf r}')U_{BB}
\hat{\Phi}({\bf r}')\hat{\Phi}({\bf r}),
\nonumber
\end{equation}
\begin{equation}
\hat{W}_{BF}=
\int \int d{\bf r} d{\bf r}' 
\hat{\Phi}^\dagger({\bf r}) \hat{\Psi}^\dagger({\bf r}')
U_{BF}\hat{\Psi}({\bf r}')\hat{\Phi}({\bf r}).
\end{equation}
\noindent 
Here $\hat{T}_{B}$ and $\hat{T}_{F}$ denote the boson
and fermion kinetic energies, 
$V_{B}({\bf r})$ and $V_{F}({\bf r})$ 
the boson and fermion trapping
potentials, and $\hat{\Phi}({\bf r})$ 
and $\hat{\Psi}({\bf r})$
the boson and fermion field operators.

Let the ground state of the system 
be $|g \rangle$, and introduce the ground-state energy
$E_0\stackrel{\mathrm{def}}{=}\langle g|\hat{H}|g \rangle$,
and the boson and fermion densities 
$n_B({\bf r})\stackrel{\mathrm{def}}{=} \langle 
g|\hat{\Phi}^\dagger({\bf r})\hat{\Phi}({\bf r})|g \rangle$,
$n_F({\bf r})\stackrel{\mathrm{def}}{=} \langle 
g|\hat{\Psi}^\dagger({\bf r})\hat{\Psi}({\bf r})|g \rangle$.
The Hohenberg-Kohn theorem~\cite{DreGro90} guarantees that,
given the interaction potentials, the ground-state energy
depends only on the densities, i.e. is a functional 
$E_0=E_0[n_B,n_F]$. The theorem was proved originally
for Fermi systems, but its generalization to Bose systems
and to Bose-Fermi mixtures is straightforward.
Determination of the density 
distributions follows by imposing the stationarity
conditions:
\begin{equation}
\frac{\delta E_0[n_B,n_F]}{\delta n_B({\bf r})}
\stackrel{!}{=}\mu_{F};
\; \; \; \;
\frac{\delta E_0[n_B,n_F]}{\delta n_F({\bf r})}
\stackrel{!}{=}\mu_{B} \, ,
\label{normalization}
\end{equation}
where $\mu_{B}$ and $\mu_{F}$ are the boson and fermion
chemical potentials.

In general, the functional $E_0[n_B,n_F]$ cannot be 
determined exactly, but we can follow the Kohn-Sham 
procedure~\cite{DreGro90} to introduce
accurate approximations. 
The idea is to map the
interacting systems of interest to a non-interacting 
reference system with the same density distributions:
$n_B({\bf r})\mapsto 
n^{\rm ref}_B({\bf r})\stackrel{!}{=}n_B({\bf r});\;
n_F({\bf r})\mapsto n^{\rm ref}_F({\bf r})\stackrel{!}{=}
n_F({\bf r}).$
Uniqueness of the mapping follows from the
Hohenberg-Kohn theorem, and we find:
\begin{eqnarray}
E_0
& = & T_{B}^{\rm ref}[n_B,n_F]+T_{F}^{\rm ref}[n_B,n_F]
\nonumber\\
&& \nonumber \\
& + & \int d{\bf r}V_B n_B 
+\int d{\bf r}V_F n_F +\frac{g_{BB}}{2}\int d{\bf r}
n_B^{2}
\nonumber\\
& + & g_{BF}\int d{\bf r} 
n_B n_F \; + E_{\rm xc}[n_B,n_F] \, ,
\label{Efinal}
\end{eqnarray}
where the first two terms are 
the kinetic energies 
of the reference system, the next two terms are the 
trapping energies, and the fifth and sixth term 
are the mean-field part of the interaction energy.
The last term includes all the contributions to the 
interaction energy beyond mean field due to exchange
correlations, and defines the exchange-correlation 
energy functional
$E_{\rm xc}[n_B,n_F]$. 
If $E_{\rm xc}$ is neglected altogether,
one simply recovers the equations of 
mean-field theory  
for trapped Bose-Fermi mixtures \cite{Mol98,Roth}. 

We now proceed to carry out the full Kohn-Sham scheme
to determine the ground-state energy, and the boson
and fermion density profiles beyond mean field.
In the Kohn-Sham reference system the kinetic 
parts of the energy functional  
$T_{B}^{\rm ref}[n_B,n_F]$ for the bosons and
$T_{F}^{\rm ref}[n_B,n_F]$ for the fermions
are defined as:   
\begin{eqnarray}
T_{B}^{\rm ref}[n_B,n_F]&=&
-N_B\int d^3{\bf r}\phi^{*}({\bf r})
\frac{\hbar^2{\bf \nabla}^2}{2m_B}\phi ({\bf r}),
\nonumber\\
T_{F}^{\rm ref}[n_B,n_F]&=&
-\sum_{i=1}^{N_F} \int d^3{\bf r}\psi_{i}^{*}({\bf r})
\frac{\hbar^2{\bf \nabla}^2}{2m_F}\psi_{i} ({\bf r}),
\label{Tnonint}
\end{eqnarray}
\noindent where $N_{B}$ and $N_{F}$ are
the total numbers of bosons and fermions,
and the notation 
$\phi({\bf r})$, $\psi_{i}({\bf r})$ 
is a shorthand for
the boson and fermion functional orbitals
$\phi[n_B,n_F]({\bf r})$ and
$\psi_{i}[n_B,n_F]({\bf r})$ of the 
non-interacting reference system. 
Inserting Eqns. (\ref{Tnonint}) 
into Eqn. (\ref{Efinal}) and carrying out
the functional derivatives in Eqns. (\ref{normalization}),
we obtain a system of coupled,
effective Schr\"odinger equations for the 
single-particle states that are the desired 
Kohn-Sham equations for a Bose-Fermi system:
\begin{eqnarray}
\left[ -\frac{\hbar^{2}{\bf \nabla}^{2}}{2m_B}
         +V_B
         +\frac{4\pi\hbar^{2}a_{BB}}{m_{B}}n_B
+\frac{2\pi\hbar^{2}a_{BF}}{m_{R}}n_F \right. \nonumber\\
\left. +\frac{\delta E_{\rm xc}}{\delta n_B}
\right] \phi = \mu_{B} \phi ,
\nonumber\\
\left[ -\frac{\hbar^2{\bf \nabla}^2}{2m_F} 
         +V_F
         +\frac{2\pi\hbar^{2}a_{BF}}{m_{R}}n_B
         +\frac{\delta E_{\rm xc}}{\delta n_F}
         \right] \psi_{\it i}
=\epsilon_{\it i} \psi_{\it i} ,
\label{KSequations}
\end{eqnarray}
with $n_B({\bf r})=N_B|\phi({\bf r})|^2$,
$n_F({\bf r})=\sum^{N_F}_{i=1}|\psi_{\it i}({\bf r})|^2$,
where the sum in $n_F({\bf r})$ runs over 
the $N_F$ single-particle
states $\psi_{\it i}$ with 
lowest energies $\epsilon_{\it i}$.
We now resort 
to local density approximation (LDA) by
approximating $E_{\rm xc}$ with an integral over the 
exchange-correlation energy density 
$E_{\rm xc}^{hom}(n_B({\bf r}),n_F({\bf r}))$
of a homogeneous system taken at the -yet unknown- 
densities $n_B({\bf r})$ and $n_F({\bf r})$:
\begin{equation}
E_{\rm xc}[n_B,n_F]\approx\int d {\bf r}
E_{\rm xc}^{hom}(n_B , n_F ).
\end{equation}
With this identification,
functional derivatives become
ordinary partial derivatives: 
\begin{eqnarray}
\frac{\delta E_{\rm xc}}{\delta n_B}=
\frac{\partial E_{\rm xc}^{hom}}{\partial n_B};
\quad \frac{\delta E_{\rm xc}}{\delta n_F}=
\frac{\partial E_{\rm xc}^{hom}}{\partial n_F} \, .
\end{eqnarray}
The homogeneous exchange-correlation energy density 
$E_{\rm xc}^{hom}$ has been recently
determined~\cite{Noi} to second order in 
the boson-fermion scattering length $a_{BF}$
via a $T$-matrix approach analog of the Beliaev
expansion for a pure Bose system~\cite{Bel58}, and
its expression reads~\cite{Noi}: 
\begin{equation}
E_{\rm xc}^{hom}\left( n_B,n_F\right) = 
\frac{2\hbar^2a_{BF}^{2}}{m_{R}}f(\delta)k_Fn_Fn_B,
\label{exc}
\end{equation}
where $k_F=(6\pi^2 n_F)^{1/3}$ is the Fermi 
wave vector, and 
$f(\delta)$ is a dimensionless function
that depends only on the boson and fermion masses: 
\begin{equation}
f(\delta) = 
1-\frac{3+\delta}{4\delta}
+\frac{3(1+\delta)^{2}(1-\delta)}{8\delta^{2}}\ln
\frac{1+\delta}{1-\delta} \, ,
\label{FM}
\end{equation}
with $\delta=(m_B-m_F)/(m_B+m_F)$.
Viverit and Giorgini have recently shown~\cite{Giorgini}
that (\ref{exc}) is exact in the
limit $k_F\xi_B \gg 1$, where
$\xi_B = 1/\sqrt{8\pi n_Ba_{BB}}$ is 
the boson healing length. In order of magnitude,
the homogeneous densities are 
$n_F\approx N_F/\ell^3$ and 
$n_B\approx N_B/\ell^3$,
where $\ell$ is the
characteristic length of the 
confining potential. The condition
$k_F\xi_B \gg 1$ is then equivalent to 
$N_F\gg N_B^{3/2}(a_{BB}/\ell)^{3/2}$.
On the other hand, LDA is correct for large
$N_{B}$ and $N_{F}$, provided that
$\ell \gg a_{BB}, a_{BF}$, i.e. that the 
characteristic lengths of the confining
potentials are much larger than the 
scattering lengths. In current experiments 
$N_{F} \approx N_{B} \approx 10^{4}$ and
$a_{BF}/\ell \approx a_{BB}/\ell \approx 10^{-3}$, 
so that the condition  $k_F\xi_B \gg 1$
is well satisfied.
Moreover, the boson-boson exchange-correlation 
energy is $256\hbar^2a_{BB}n^2_B\sqrt{\pi 
n_Ba^3_{BB}}/15m_B$ (see e.g. \cite{Bel58}).
This is much smaller than the 
exchange-correlation energy (\ref{exc})
if $N_F>>5.4(a_{BB}/a_{BF})^{3/2}(a_{BB}/\ell)^{3/8}((1
-\delta)/f(\delta))N_B^{9/8}$. 
Since $a_{BB}/a_{BF}=0.13$ for the 
Paris experiment with $^{6}$Li-$^{7}$Li
\cite{SchKha01} and $a_{BB}/a_{BF}=0.28$ for
the Florence experiment with $^{40}$K-$^{87}$Rb
\cite{RoaRib02} (these are the only two
experiments where $a_{BF}$ has been measured),
this condition is satisfied as well.
Yet other higher order terms are due to
direct Fermion--Fermion $p$-wave scattering. These
terms are at least of order $(k_{F}a_{FF})^3$, 
where $a_{FF}$ is the Fermion--Fermion $p$--wave 
scattering length, and thus certainly
negligible against the term we consider.
Altogether, Eq. (\ref{exc}) provides the 
most relevant contribution to 
the exchange--correlation energy 
for the current experimental situations.
For more general situations, Eq. (\ref{exc})
provides the most relevant contribution
beyond mean field any time LDA is satisfied,
$N_{F}$ is comparable or larger than $N_{B}$
in order of magnitude, and perturbation theory 
holds, i.e. $k_{F}a_{BF}/\pi << 1$, and a
sufficiently small Bose gas parameter. 

We now consider
the Kohn--Sham system 
(\ref{KSequations}) with the exchange-correlation
energy (\ref{exc}) for spherically symmetric, 
harmonically trapped systems:
$V_B({\bf r})=(m_B\omega_B^2r^2)/2$,
$V_F({\bf r})=(m_F\omega_F^2r^2)/2$.
Due to the spherical symmetry we 
can write:
\begin{eqnarray}
\phi({\bf r})=\frac{u(r)}{r}Y_{00}; \quad
\psi_{nlm}({\bf r})=\frac{u_{nl}(r)}{r}Y_{lm},
\end{eqnarray}
where $Y_{lm}(\Theta,\Phi)$ 
are the spherical harmonics, and the 
Kohn-Sham equations (\ref{KSequations})
become:
\begin{equation}
\left[ -\frac{1}{2m_B}\frac{d^2}{dr^2}
        +\frac{m_{B}}{2}\omega_B^2r^2
	+\frac{4\pi a_{BB}}{m_B}n_B(r)
	+\frac{2\pi a_{BF}}{m_R}n_F(r)
        \right. 
\nonumber
\end{equation}
\begin{equation}
\left. +\frac{2 a^2_{BF}f(\delta)}{m_R}n_F(r)k_F(r)
         \right] u(r) =\mu_{B}u(r) \, ;
\nonumber
\end{equation}
\begin{equation}
\left[ -\frac{1}{2m_F}\frac{d^2}{dr^2}
        +\frac{l(l+1)}{2m_Fr^2}
        +\frac{m_{F}}{2}\omega_F^2r^2
	+\frac{2\pi a_{BF}}{m_R}n_B(r)
	\right. 
\nonumber
\end{equation}
\begin{equation}
\left.  +\frac{8a^2_{BF}f(\delta)}{3m_R}n_B(r)k_F(r)
         \right] u_{nl}(r) =\epsilon_{nl}u_{nl}(r) \, ,
\label{KSE}
\end{equation}
with $\int dr\; u^2(r)=1,\int dr\; 
u^2_{nl}(r)=1$, where
$n$ denotes the number of nodes of the
radial functions $u_{nl}$. 
The normalized
density distributions $\tilde{n}_{B}({\bf r})
=4\pi r^{2}n_{B}({\bf r})$ and
$\tilde{n}_{F}({\bf r})=4\pi r^{2}n_{F}({\bf r})$ 
are:
\begin{equation}
\tilde{n}_{B}({\bf r}) = N_{B} u^2(r) \, , 
\end{equation}
and
\begin{equation}
\tilde{n}_{F}({\bf r}) =
\sum_{\epsilon_{nl}\leq\mu_F}(2l+1)u^2_{nl}(r) \, .
\end{equation}

\section{Solution of the Kohn--Sham equations}
\label{Results}
The above expressions together with
Eqns. (\ref{KSE})
define a system of coupled nonlinear
differential equations. 
The numerical solution is obtained iteratively. 
We initialize $n_B({\bf r})$ and $n_F({\bf r})$ 
to be the Thomas--Fermi density distributions
with no boson--fermion coupling. 
We then use these as initial densities for 
Eqns. (\ref{KSE}). The energy eigenvalues 
are found by a bi-section algorithm, iterating
the procedure to the desired degree of accuracy.
Knowing the states $u$ and $u_{nl}$, one
must determine the wave function 
$u_{nl}$ with lowest energy $\epsilon_{nl}$ 
using the fact that 
$\epsilon_{nl}$ grows with $n$ and $l$.
When all the occupied Kohn-Sham states 
are determined, the output densities are 
compared to the 
initial distributions. If they are about 
the same, a self-consistent solution is reached, 
and the procedure ends. If not, one defines a convex
combination of the initial and output densities,  
$n^{new}_{B(F)}(r)=(1-x)\cdot n_{B(F)}^{initial}
+x\cdot n_{B(F)}^{output}$, with $0 < x\leq 1$,
and iterates the procedure until
convergence is reached with the desired degree
of accuracy. 
If $N_{F}$ is large, the procedure is very time
consuming and limited by a maximum number of nodes
that can be included.  
One then adopts a Thomas-Fermi 
approximation for the fermion kinetic energy,
whenever $N_F \geq 1000$, and finds a posteriori
a very good agreement with the single-particle
description.

Comparison of our results with current experiments 
can be carried out for those systems whose
boson-fermion scattering length has been measured.
These are the $^6$Li-$^7$Li mixture realized in the 
Paris experiment~\cite{SchKha01}, and the
$^{40}$K-$^{87}$Rb recently realized in the Florence
experiment~\cite{RoaRib02}.
In the Paris experiment with fermionic $^6$Li
and bosonic $^7$Li, the
measured scattering lengths 
are $a_{BB}=5.1a_0$ and $a_{BF}=38.0a_0$, where
$a_0$ is the Bohr radius. Taking $\omega_{B}$ as the
unit of frequency, the exchange-correlation 
energy turns out to be $\approx 50\hbar\omega_{B}$, whereas the
mean-field boson-fermion interaction 
energy is $\approx 7455\hbar\omega_{B}$. 
Thus only about $0.67\%$ of the interaction energy is
due to exchange correlations, it has the same
sign of the mean-field energy, and the modification
of the mean-field density profiles is negligible.

\begin{figure}[ht]
\begin{center}
\epsfysize=8.5cm
\rotatebox{-90}{\epsfbox{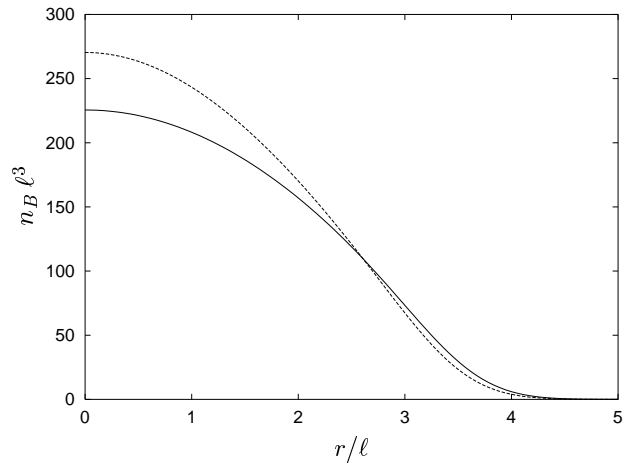}}
\end{center}
\caption{The boson density profile for the Florence experiment.
Dashed line: without exchange correlations;
solid line: with exchange correlations. Quantities
are dimensionless, rescaled in units of
$\ell = (\hbar/m_{B}\omega_{B})^{1/2}$.}
\label{Rbden}
\end{figure}

\begin{figure}[ht]
\begin{center}
\epsfysize=8.5cm
\rotatebox{-90}{\epsfbox{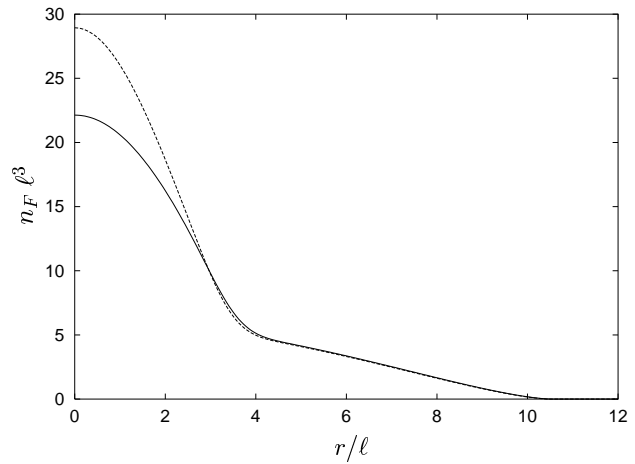}}
\end{center}
\caption{The fermion density profile for the Florence experiment.
Dashed line: without exchange correlations;
solid line: with exchange correlations. Quantities
are dimensionless, rescaled in units of 
$\ell = (\hbar/m_{B}\omega_{B})^{1/2}$.}
\label{Kden}
\end{figure}

The situation is very different for the mixture
of fermionic $^{40}$K and bosonic $^{87}$Rb 
realized in the Florence experiment, due to the large
and negative boson-fermion
scattering length giving rise both to a large
attractive mean--field boson--fermion interaction
potential, and to a non negligible 
exchange--correlation potential. The latter,
being proportional to the square of the
boson-fermion scattering length, is always
repulsive.
For this experiment, a typical stable configuration
is achieved for $N_F = 10^{4}$,
$N_B = 2\times 10^{4}$. The boson--boson scattering
length is $a_{BB} = 100a_0$, while the boson--fermion
scattering length $a_{BF} \approx -400a_0$ is
measured with an uncertainty of about $50\%$.
The mean-field interaction 
energy is $\approx  -98165\hbar\omega_{B}$,
while the exchange-correlation energy is 
$\approx 6783\hbar\omega_{B}$. 
Thus the relative correction in the interaction
energy is about $7\%$ of the mean-field result,
going in opposite direction, and leads to a  
pronounced effect on the density profiles. 
Both the boson and fermion densities
spread out and decrease substantially at the center of the trap 
with respect to the mean-field prediction, due to the
repulsive exchange--correlation potential.
This effect is shown in
Figs.~\ref{Rbden} and \ref{Kden},
where we show the boson and fermion density distributions 
with and without exchange correlations, calculated with
the parameters fixed at the values measured in
the Florence experiment. 
At the center of the trap the
boson and fermion densities are reduced,
respectively, to about $85\%$ and $78\%$ 
of the mean-field result.  
 
\section{Stability and Collapse}
\label{Stability}
In general there are two kinds of instabilities in a binary mixture 
(we do not consider instabilities due to fermion pairing):
demixing \cite{Mol98} and simultaneous collapse of both
the boson and the fermion component \cite{Roth02}. 
The first can occur if the interaction between the two species
is repulsive, and implies by definition a minimal overlap of the 
density distributions. In this
case we do not expect a significant change of the
phase diagram by repulsive exchange--correlation
interactions, but only for a small enhancement of the 
phase separation. 

In the collapse regime, which can occur if the
interaction between the two species is attractive, 
the situation is radically different, as in this
case one has indeed a very high overlap of the densities
in the center of the trap. 
The exchange--correlation interaction, which is always
repulsive to second order in the boson-fermion scattering
length, opposes the propensity to collapse due to
the attractive mean-field contribution. If the
coupling strength between the two components 
of the mixture is sufficiently strong, the
exchange--correlation can significantly modify
the phase diagram.

\begin{figure}
\begin{center}
\epsfysize=8.5cm
\rotatebox{-90}{\epsfbox{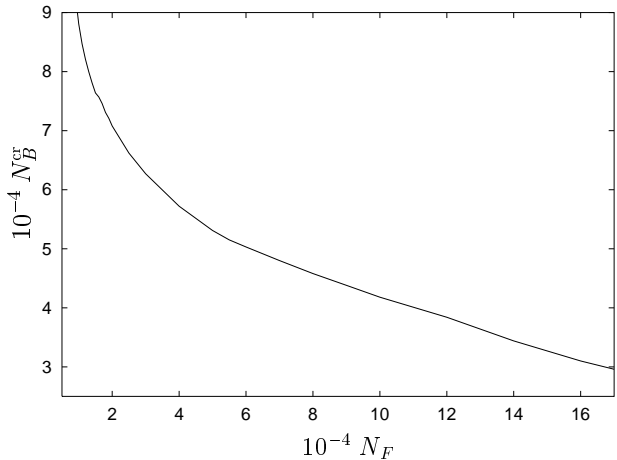}}
\end{center}
\caption{The critical number of bosons $N_{B}^{cr}$ 
for the onset of collapse as a function of the
number of fermions $N_{F}$ in mean--field approximation.}
\label{ncrmf}
\end{figure}

\begin{figure}
\begin{center}
\epsfysize=8.5cm
\rotatebox{-90}{\epsfbox{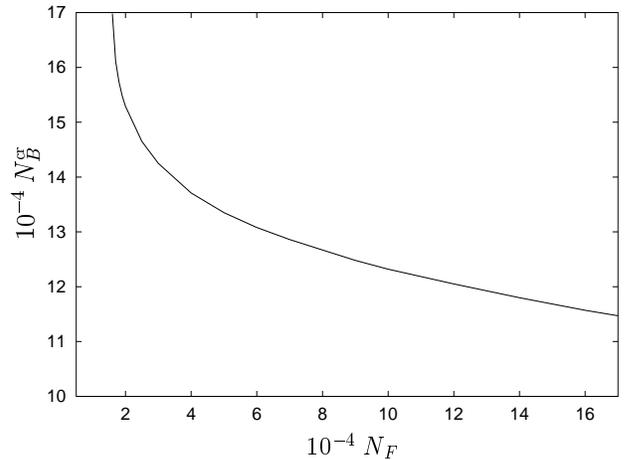}}
\end{center}
\caption{The critical number of bosons $N_{B}^{cr}$ 
for the onset of collapse as a function of the
number of fermions $N_{F}$ including 
exchange--correlation.}
\label{ncrxc}
\end{figure}

In Fig. \ref{ncrmf} we provide the mean--field
phase diagram of a binary boson--fermion mixture,
with the physical parameters of the 
Florence experiment \cite{ModRoa02}. The plot
shows the behavior of the critical number
of bosons $N_{B}^{cr}$, i.e. the threshold number 
for the onset of collapse , as a function of the 
number of fermions $N_{F}$. Collapse occurs
at any point of the phase plane above the critical
curve, while the mixture is stable at all points
below it. For low fermion numbers $N_{F} \leq 8 
\times 10^{3}$, the critical number of bosons 
$N_{B}^{cr}$ begins to grow so fast that to all 
practical purposes collapse is inhibited. 
The inversion regime between the number of fermions
and the critical number of bosons takes place
at $N_{F} \simeq N_{B}^{cr} \simeq 5 \times 10^{4}$.
For a typical number of fermions $N_{F} \simeq 2 \times
10^{4}$ one has a critical boson number
$N_{B}^{cr} \simeq 7 \times 10^{4}$.
The situation in mean-field approximation is to
be compared with the prediction obtained by
including exchange--correlation.
Fig. \ref{ncrxc} shows the same phase diagram 
as in Fig. \ref{ncrmf} but with the inclusion
of exchange--correlation. 
We clearly see a significant increase in the critical 
number of the bosons due to exchange--correlation. 
The inversion regime between the number of fermions
and the critical number of bosons takes place at
$N_{F} \simeq N_{B}^{cr} \simeq 1.2 \times 10^{5}$,
and for a typical fermion number 
$N_{F} \simeq 2 \times 10^{4}$ the critical boson number
$N_{B}^{cr} \simeq 1.5 \times 10^{5}$, i.e. a much
larger number of bosons is needed to produce a collapse
of the fermion component.
This behavior was qualitatively expected
since the effective exchange-correlation potentials are 
always repulsive to second order in the boson-fermion
scattering length. 

The quantitative difference between the mean--field and
the exchange--correlation phase diagrams deserves
some explanatory comments.
First of all, the determination of the critical line for
simultaneous collapse takes place in a regime where the 
numerics is very sensitive to small deviations of the
input parameters.
Thus, when a stable solution is not found, this could be
ascribed either to the fact that the physical collape regime 
was reached or to an inappropriate numerical precision. 
However, by increasing the numerical precision, 
computation time rapidly increases as well.
On the other hand, if a stable numerical solution is found, 
there can certainly be no physical collapse. 
The critical curves we present
are then lower bounds on the critical numbers.
We remark that the mixture is very sensitive
to the exact value of the boson--fermion scattering length 
in the collapse regime.
Since this value is experimentally known with a large
uncertainty, it would be crucial to determine it with a much
greater precision. This could be achieved by tuning the
scattering length in order to fit the experimental data 
on the onset of collapse \cite{Modugno}.
Moreover, for large interaction strengths, such as that in the Florence 
experiment, the second--order term in the exchange--correlation 
energy might overestimate the effect of stabilization. 
In fact, in these cases, the attractive third--order term could
possibly give rise to a non negligible contribution, 
so that the mean--field critical line of Fig. \ref{ncrmf} 
and the second--order critical
line of Fig. \ref{ncrxc} would provide, respectively, a lower
and an upper bound. The true phase--diagram would therefore
lie in between the two. A more detailed analysis than
that provided in the present paper requires however
analytical expressions of the third--order interaction energy in 
powers of $k_{F}a_{BF}$, and this is a formidable task,
because Feynman diagrams containing all
possible combinations of Boson--Fermion and Boson--Boson interactions
have to be considered. These effects cannot be simply 
determined by resumming restricted classes of equivalent
diagrams. Finally, to go beyond second--order perturbation
theory requires, for consistency, to take into account interaction 
processes beyond $s$--wave, such as $p$--wave scatterings, thus 
introducing powers of, e.g., the $p$--wave Boson--Fermion
scatteting length, and the description soon
becomes exceedingly complex in the framework of perturbation
theory. Non perturbative methods like Monte Carlo 
simulations would then be desirable to establish more
accurate results.

In conclusion, we have introduced
the Kohn-Sham scheme of DFT for inhomogeneous
Bose-Fermi systems to determine the ground-state
energy and density profiles to second--order
in the boson--fermion scattering length.
We have compared the theoretical predictions with 
current experiments, discussed the relavance of
different exchange--correlation terms, and
investigeted the 
the importance
of the exchange-correlation effects for dilute atomic gases. We have shown
that they are
substantial for systems, like $^{40}$K-$^{87}$Rb,
with a large attractive boson-fermion interaction,
especially in the critical regime of collapse onset,
by comparing the mean--field and the exchange--correlation
phase diagrams.
The DFT method outlined here can be in principle extended to
include higher--order corrections and finite temperature
effects. 

We thank H. Hu and M. Modugno for useful 
comments on an earlier draft of our work,
and for stimulating conversations.
A. A. and M. W. thank the DFG and the ESF
for financial support.
F. I. thanks the INFM for financial support.

\end{document}